\documentclass[11pt]{article}
\usepackage{epsf,amsmath,amssymb}
\begin{document}

\begin{flushright}
 SU-HE 686 \\
 BI-TP-98/27 \\
 September 1998
\end{flushright}

\begin{center}

\vspace{24pt}
{\Large \bf Monte Carlo Renormalization of 2d Simplicial  \\
 \vspace{5pt} 
 Quantum Gravity Coupled to Gaussian Matter}
\vspace{24pt}

{\large \sl E.B.~Gregory $^a$,  S.M.~Catterall $^a$ \,{\rm and}\, 
  G.~Thorleifsson $^b$ } \\
\vspace{10pt}
$^a$ Physics Department, Syracuse University, Syracuse, NY 132144, USA \\
$^b$ Facult\"{a}t f\"{u}r Physik, Universit\"{a}t Bielefeld
     D-33615, Bielefeld, Germany  \\
\vspace{10pt}     

\begin{abstract}
We extend a recently proposed real-space renormalization
group scheme for dynamical triangulations to situations
where the lattice is coupled to continuous scalar fields. Using
Monte Carlo simulations in combination with a linear, stochastic
blocking scheme for the scalar fields we are able to determine the
leading eigenvalues of the stability matrix with good accuracy both 
for $c_M=1$ and $c_M=10$ theories.  
\end{abstract}

\end{center}
\vspace{15pt}

\section{Introduction}

Dynamical triangulations offer a powerful way to investigate two-dimensional 
quantum gravity and, when coupled to scalar fields, bosonic string theories.
This manner of discretizing such theories allows the use of techniques
drawn from statistical mechanics. Great strides have been made in
analytically understanding such models, notably with the development of
matrix model techniques for solving the discrete systems
and continuum approaches
based on quantizing Liouville theory \cite{KPZDDKref}.  
However, all these analytic approaches break down when
applied to systems whose matter central charge $c_M$ is greater
than unity. Furthermore, the analytic approaches  
allow us to compute only correlation functions of {\it integrated} 
operators --- they
yield very little information on the nature of the quantum geometry
or physically interesting questions concerning matter field correlators
defined on geodesic paths.  

To answer such questions, we must rely on 
Monte Carlo simulation techniques.  Combining this powerful
numerical tool with another, the renormalization group, in the
form of the Monte Carlo Renormalization Group (MCRG), provides new
insight into the critical behavior and hence the continuum limit for
these models.

In this paper we describe our efforts to develop an efficient and
functional blocking scheme for models of continuous scalar
fields coupled to 2d dynamical triangulations. 
We first introduce the class of model we consider and describe our
simulation algorithm. Then we discuss real-space renormalization
groups (RG) and the particular blocking schemes we have investigated. 
The primary goal of this work has been to extend the MCRG techniques, 
developed in Ref.~\cite{ThorCatt96}, to continuous scalar fields.
As a test of this method we have investigated models of one ($c_M = 1$) 
and ten ($c_M = 10$) scalar fields coupled to two-dimensional
gravity respectively.  The former model is chosen as it is
solvable in the continuum, the latter as it represents a system 
where the back-reaction of matter is so strong that the internal geometry 
degenerates into polymer-like structure.
As this RG method involves blocking a dynamical geometry, we start by
demonstrating that it preserves the relevant fractal structure 
of the triangulations, characterized by such geometric exponents as 
the Hausdorff dimension $d_H$ and the string susceptibility 
exponent $\gamma_s$.  For the matter sector we have applied
a linear stochastic blocking scheme to the fields and, for
both the models we consider, after optimizing the blocking
procedure, we determine the critical exponents governing the
rescaling of the fields. 

\section{Model and Numerical Approach}

The model we examine is a two-dimensional dynamically triangulated surface 
coupled to $D$ copies of Gaussian scalar fields.  It has a canonical 
(fixed area) partition function
\begin{equation}
 Z \;=\; \sum_{\tau,\phi} \; {\rm e}^{\textstyle - S_\tau [\phi]},
\label{partition}
\end{equation}
where the sum is over all possible combinatorial 
triangulations\footnote{A priori, different ensembles of triangulations
can be used provided they lead to a well defined partition function
Eq.~(\ref{partition}).  In this work we use combinatorial 
triangulations, excluding self-energy and tadpole diagrams from
the corresponding dual graph, as this simplifies the geometric blocking.} 
$\tau$, as well as over all possible configurations of the fields 
$\phi$.  The action, 
\begin{equation}
 \label{action}
 S[\phi] \;=\; \sum_{\mu=1}^D \; \sum_{<i,j>} \left 
 (\phi_i^\mu - \phi_j^\mu \right )^2,
\end{equation}
depends on $\mu = 1,...,D \;\; (= c_M)$ scalar fields $\phi^{\mu}$, 
and the second sum is over all nearest 
neighbor lattice points, $i$ and $j$.  Notice that due to the
back-reaction of the matter fields on the geometry,
coupling multiple copies of scalar fields to dynamical
triangulations leads to genuinely different statistical systems, 
as the partition functions do not factorize, in contrast
to the behavior on a regular lattice. 
In this paper we investigate models with one and ten fields, 
corresponding to matter with central charge $c_M=1$ and 10
respectively.

The partition function Eq.~(\ref{partition}) is evaluated numerical
using Monte Carlo methods.  The triangulations are update 
using a standard link-flip move --- a link $l_{ab}$ connecting
two adjacent triangles $t_{abc}$ and $t_{bad}$ is removed
and replaced by the link $l_{cd}$.
We accept or reject such a move based on a Metropolis test.  

\begin{figure}
 \epsfxsize=4in \centerline{\epsfbox{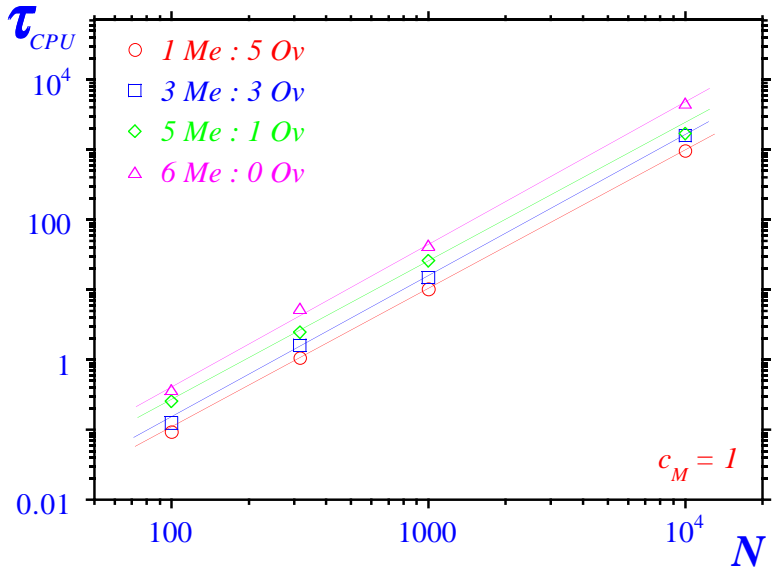}}
 \caption{\small The auto-correlation time $\tau_{CPU}$, measured
  in real-time, for different volumes $N$ and for different ratios 
  of Metropolis {\it versus} overrelaxation updates.  
  The auto-correlations are estimated for the slowest mode, 
  the radius of gyration $\sum_i \phi_i^2$,
  for the model Eq.~(\ref{partition}) coupled to one scale field.} 
 \label{ovr_rlx}
\end{figure}

Updates of the Gaussian fields proceed in two ways. The first is a 
standard Metropolis update, where a small (local) change in
the fields is accepted/rejected based on a Metropolis test.  
However, as this updating procedure suffers from very long 
auto-correlations we also employ an overrelaxation algorithm to 
update the fields. This involves replacing one of the fields 
at a node $i$ with
\begin{equation}
 \phi_i \;\rightarrow\; \phi_i' \;=\; \left [\frac{2}{q_i}
 \sum_{<i,j>} \phi_j\right ]-\phi_i \; ,
 \label{ovrrelx}
\end{equation}
where $j$ indexes the $q_i$ neighbors of  $i$. Since
$ \sum_{<i,j>}(\phi_i-\phi_j)^2=\sum_{<i,j>}(\phi_i'-\phi_j)^2$,
the action is preserved and the move is automatically accepted.
The overrelaxation algorithm is, however, non-ergodic
and some amount of Metropolis updates have to be included.  
We find, nonetheless, that by using a ratio of only 
one Metropolis update to every five overrelaxation updates 
the auto-correlations, measured in real-time, are reduced by 
about a factor of four compared to a pure Metropolis update. 
This is illustrated in Figure \ref{ovr_rlx}. 
This reduction can be understood qualitatively as the overrelaxation
suppresses the usual random walk behavior of local updating
algorithms \cite{neal}.  
However, in contrast to overrelaxation applied to scalar
fields on a regular lattice \cite{kennedy}, on dynamical
triangulations critical slowing down is not reduced,
only the overall prefactor  --- the dynamics of
the updating procedure are dominated 
by the local geometric moves.

\section{Monte Carlo Renormalization}

A powerful tool for investigating the critical behavior of statistical 
systems is the renormalization group.  The usual approach is to
use some ``course graining'' procedure whereby the system is blocked
or replaced by one with fewer degrees of freedom, effectively 
integrating out the short distance fluctuations while preserving the 
long-distance physics of the system.
This renormalization group operation can be written in 
terms of a projection operator, ${\cal P}$, that maps the 
system from the old degrees of freedom to the new ones while 
preserving the partition function Eq.~(\ref{partition}):
\begin{equation}
 \label{transform}
 {\rm e}^{\; S'\left[\phi' \right]}
 \;=\; \sum_{\bf \phi} {\cal P}\left [{\bf \phi'},{\bf \phi}\right]
 {\rm e}^{  S\left[\phi \right]}, 
 \qquad {\rm with} \qquad
 \sum_{\phi^\prime} {\cal P}\left[\phi^\prime,\phi\right] \;=\; 1.
\end{equation}

In general, upon application of the renormalization group the action
$S[\phi]$, expanded on a suitable basis of operators 
$\{ {\cal O}_{\alpha} \}$, changes according to
\begin{equation}
 S \;=\; \sum_\alpha K_\alpha {\cal O}_\alpha 
 \;\;\stackrel{\cal P}{\longrightarrow}\;\;
 S' \;=\; \sum_\alpha K_\alpha' {\cal O}_\alpha' \,.
\end{equation}
The $K$'s and the $\cal{O}$'s are coupling constants and operators
defined on the bare lattice, while their primed counterparts
denote the corresponding quantities on the blocked lattice. 
Repeated iteration of the RG transformation
leads to a flow in the associated coupling constants towards a
fixed point $\{K^*\}$:
\begin{equation}
 {\{K\}}^{(0)} \;\stackrel{\cal P}{\rightarrow}\; {\{K\}}^{(1)} 
 \;\stackrel{\cal P}{\rightarrow}\; ... \;\stackrel{P}{\rightarrow}\; 
 {\{K^*\}} \,.
\end{equation}
A linearized approximation to the RG transformation
in the vicinity of a critical (unstable) fixed point:
\begin{equation}
 \label{expand2}
 \delta K_\alpha^{(k+1)} \;=\; K_\alpha^{(k+1)}-K_\alpha^*
 \;\simeq\; 
 \sum_\beta \left. \frac{\partial K_\alpha^{(k+1)}}
 {\partial K_\beta^{(k)}} \right|_{\bf K=K^*} \hspace{-15pt} \delta K_\beta 
 \;\equiv\; \sum_\beta T_{\alpha \beta} \, \delta K_\beta^{(k)},
\end{equation}
yields an eigenvalue equation for the stability
matrix $T_{\alpha \beta}$,
\begin{equation}
 \sum_\beta T_{\alpha \beta} \, u^i_\beta \;=\; 
 \lambda_i \, u^i_\alpha.
\end{equation}
An eigenvalue $\lambda_i$, corresponding to a relevant
operator $u^i_{\alpha}$ in the effective action, 
defines a critical exponent $y_i$ associated with the 
fixed point:  $\lambda_i = b^{\,y_i}$,
where $b=N^{(k)}/N^{(k+1)}$ is the volume blocking factor.

For some simple systems one may be able to write down an exact 
expression for the projection operator ${\cal P}$ and
determine the critical properties explicitly from the stability matrix. 
In general, though, this is not possible.
For dynamical triangulations, where the blocking involves
both the geometry and the fields living on the surface, this
task is even more complicated.
Especially since, as discussed in next section, an explicit form of a
projection operator for blocking the geometry is not available.
For these reason we resort to a Monte Carlo 
Renormalization Group procedure.  This allows the determination of the
elements of the stability matrix through Monte Carlo simulations
in combination with a local blocking method for both the fields
and the geometry:
\begin{equation}
 \frac{\partial \langle {\cal O}_\gamma^{(k+1)}\rangle}
   {\partial K_\alpha^{(k)}}
 \;=\; \sum_\alpha \frac{\partial K_\alpha^{(k+1)}}
   {\partial K_\beta^{(k)}}
 \frac{\partial \langle {\cal O}_\gamma^{(k+1)} \rangle}
   {\partial K_\alpha^{(k+1)}} \, ,
\end{equation}
where
\begin{equation}
 \frac{\partial \langle {\cal O}_\gamma^{(k^{\prime})} 
   \rangle}{\partial K_\alpha^{(k)}}
 \;=\; \langle {\cal O}_\gamma^{(k^{\prime})} \rangle 
   \langle {\cal O}_\alpha^{(k)} \rangle 
 - \langle {\cal O}_\gamma^{(k^{\prime})} 
  {\cal O}_\alpha^{(k)} \rangle \, .
\end{equation}

\subsection{\it Blocking the Geometry}

In dealing with fields on a dynamical triangulation we need a
renormalization group prescription that integrates out the small
scale feature of both the geometry and of the field configuration
while preserving the large scale physics.
The prescription we apply separates these two tasks.  We first
block the geometry independently from the fields by removing
nodes at random, then assign new blocked scalar fields to 
the renormalized lattice.  In this way the blocked
triangulation serves as an inert scaffold for the field blocking.  
This method has previously been applied to an Ising model 
coupled to 
dynamical triangulations, where it yielded surprisingly
accurate estimate of 
the critical properties of the model \cite{ThorCatt96}. 

\begin{figure}
 \epsfxsize=5in \centerline{\epsfbox{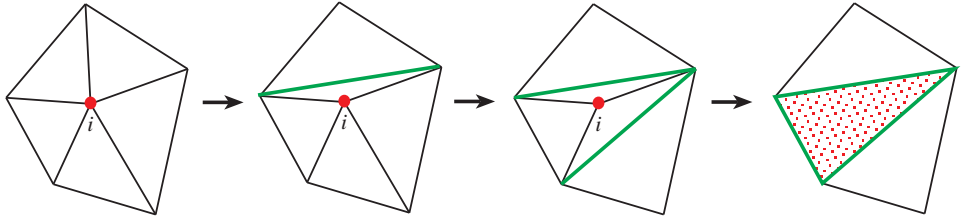}}
 \caption{\small Node decimation: A sequence of link-flips 
  around the selected node $i$ 
  reduce its coordination number to three.  Then the node is removed, 
  leaving a new triangle in its place.} 
\label{node_dec}
\end{figure}

The general idea for blocking the lattice geometry
utilizes a scheme called {\it node decimation}.  This proceeds by 
randomly picking nodes and removing them from the triangulation.
Removing a node with a coordination number $q$ will leave a 
$q$--sided hole; that hole must be randomly triangulated.  
In practice, this is done by randomly flipping links around 
the selected node until its coordination number is reduced
to three.  Now when the node is removed the three-sided hole in the 
triangulation can be replaced by a triangle.  
This is illustrated in Figure \ref{node_dec}.
This procedure is then repeated until a fraction of the
nodes, corresponding to a desired blocking factor $b$,
has been removed.  

The link-flips in the intermediate step are performed independent
of the scalar fields residing on the adjacent nodes.
This amounts to choosing the different re-triangulations of the
polygonal hole with equal probability provided no
geometric constraint is violated ---
demanding that the result is a combinatorial triangulations
may restrict the possible re-triangulations.
As this restriction is non-local, it is not possible to
write down an explicit projection operator for the
node decimation.
Alternative weight distributions for the re-triangulations
where explored in Ref.~\cite{ThorCatt96}; however, the
observed critical behavior was not very sensitive to
the particular choice.

This blocking method is justified {\it a posteriori} by 
verifying that it preserves the long distance physics, 
or the fractal structure, of the geometry.  
This we do in Section~4.1.

\subsection{\it Blocking the Scalar Fields}

For the scalar fields we have applied a {\it linear}
RG transformation to assign the block lattice fields.
For a node $i$ that survives the decimation procedure 
a new scalar field $\phi'_i$ is constructed
as a function of the original fields $\phi$ on the bare
lattice: 
\begin{equation}
 \label{fieldblock}
 \phi'_{i} \;=\; 
 \xi \left[\alpha\phi_{i}+(1-\alpha) \frac{1}{q_{i}}\sum_{j=1}^{q_i}
\phi_j \right] + \frac{\chi}{\sqrt{a_w}} \, .
\end{equation}
Here $\xi$ is an overall rescaling of the fields, and
$\alpha$ determines the relative weight between the
contribution of field on the bare lattice node $i$ and
the contribution from its bare lattice neighbors $\phi_j$.
Notice that blocking is {\it stochastic} 
--- $\chi$ is a Gaussian random noise with 
$\langle \chi^2 \rangle=1$.  The amplitude
of the noise depends on a auxiliary parameter $a_w$.

The RG transformation Eq.~(\ref{fieldblock}) can 
be expressed in terms of a projection operator
\begin{equation}
 {\cal P}\left[ \phi^\prime, \phi\right] \;=\; \exp \left[
 -\frac{a_w}{2} \sum_{i} \left (\phi_{i}'-\xi \left[\alpha\phi_i+(
 1-\alpha)\frac{1}{q_i}\sum_{j=1}^{q_i}\phi_j \right]\right )^2 \right],
\end{equation}
and depends on {\it three} parameters: $\alpha$, $\xi$ and
$a_w$.  As these parameters are crucial for a successful
blocking procedure it is necessary to establish some 
criteria for how to choose their optimal values:

\vspace{5pt}
\noindent
({\tt i}) \hspace{5pt}
The relative weight $\alpha$, between the contribution from
the fields on the initial bare lattice node and on its neighboring
nodes, must be chosen appropriate for the particular geometric 
blocking factor $b=N'/N$ being used.  
Clearly as $b\rightarrow 1$, $\alpha$ should approach one, and as 
$b$ grows $\alpha$ should decrease.  However, we have found that
in practice this scheme is fairly robust under reasonable 
choices of $\alpha$.  In this paper we us $\alpha = 0.5$.

\vspace{5pt}
\noindent
({\tt ii}) \hspace{3.5pt}
The amount of noise in the blocking procedure is controlled
by the parameter $a_w$. In general each choice of $a_w$ will lead, 
under renormalization, to a local fixed point.  This
results in a line of fixed points, all 
corresponding to the same continuum limit but  
differing in the range of their interactions on the lattice. 
For an optimal RG transformation the value of $a_w$ should
correspond to as local effective action as possible and,
in addition, should bring us in a vicinity of a fixed point 
with a minimal amount of blocking.  
In practice, the optimal value $a_w^*$ is
determined as the one yielding the most rapid convergence 
of the eigenvalues $\lambda_i$ in the blocking procedure.

\vspace{5pt}
\noindent
({\tt iii}) \hspace{1pt}
The most difficult is the determination of an appropriate value 
for $\xi$ --- the {\it field renormalization constant}.  In general, 
an arbitrary choice of $\xi$ will produce fixed point behavior, but 
only with one choice, $\xi^*$ will the action flow towards the non-trivial,
local fixed point representing the continuum limit. Therefore simply 
looking for stability of eigenvalues is not sufficient to determine  $\xi^*$.
One option is to keep the expectation value of some long-range 
observable, such as the radius of gyration $\langle \phi^2 \rangle$, 
fixed under the blocking (as done by Lang \cite{Lang86}).
We use, however, an alternative method
discussed in Ref.~\cite{BellWils75}.
It uses the property of the theory Eq.~(\ref{partition})
that an arbitrary re-scaling of the
fields does not change the physical content.
This implies that the fixed point action should include a 
marginal operator corresponding to these (redundant)
perturbations.   Associated to the marginal operator
is a sub-leading eigenvalue equal to one; requiring
that such an eigenvalue exist in the spectrum of
the stability matrix provides a criteria for choosing the
optimal value $\xi^*$.  

\vspace{5pt}
To summarize our strategy: We determine $a_w^*$ as the
value yielding the most stable eigenvalues
of the stability matrix. Using this value of $a_w$,
we then determine the correct renormalization constant
$\xi^*$ by looking for a marginal sub-leading eigenvalue.

In calculating the stability matrix in the MCRG analysis,
we have used the following basis of field operators:
\begin{equation}
 \label{basis}
 {\cal O}_m \;=\; \sum_i \phi_i \square^m\phi_i \,, 
 \qquad m=0,1,2...\,,
\end{equation}
where
\begin{equation}
 \label{box}
 \square \phi_i \;=\; \sum_{\langle ij\rangle}\phi_j-q_i\phi_i \,.
\end{equation}
Measured in lattice units, an operator
$\phi_i \square^m\phi_i$ extends $m$ steps away from the node $i$.
Hypothetically, a lattice of volume $N$ could require a basis with 
operators up to order $m \sim N^{1/d_H}$; in practice, however,
including operators with $m \leq 4$ proved to be 
sufficient.

\section{Numerical Results}

We have simulated the model Eq.~(\ref{partition}) for
one and ten scalar fields on
triangulations of up to $N = 4000$ triangles.  
A typical run consisted of about five million sweeps, 
each sweep includes approximately $N$ link-flips,
$N$ Metropolis updates, and $10N$ overrelaxation updates.
For each volume we collected and stored few thousand
independent configurations, each of which is then blocked
using a node decimation with a volume blocking factor
$b = 2$.  Storing the configurations was essential
as it allowed us to repeat the blocking and to 
analyze a wide range of values of 
the parameters $a_w$ and $\xi$. 

\subsection{\it Geometric Properties}

To demonstrate that the node decimation preserves the
long distance fractal structure of the geometry,
we have measured two different geometric exponents:
the internal {\it Hausdorff} or {\it fractal dimension} $d_H$ 
and the {\it string susceptibility exponent} $\gamma_s$.

The Hausdorff dimension provides a measure of the intrinsic
``dimensionality'' of the triangulations and is defined
by the volume of a geodesic ball with radius $r$:
$v(r) \sim r^{d_H}$.  A convenient way of measuring $d_H$
in numerical simulations is provided by the node-node
distribution function $n(r,N)$ --- the number
of nodes at a geodesic distance $r$ from a marked node.
Simple scaling arguments \cite{CattThor95,AmbJur95} imply that
\begin{equation}
 \label{ppscale}
 n(r,N) \;=\; N^{1-1/d_H} f \left(\frac{r}{N^{1/d_H}}\right),
\end{equation}
and $d_H$ is determined as the value that optimally ``collapses''
distributions $n(r,N)$, measured on different lattice volumes,
on a single scaling curve.
Notice, however, that in determining $d_H$ on blocked 
triangulations only distributions $n^{(k)}(r,N)$ 
corresponding to the same amount of node decimation $k$
should be included in the analysis.

In Table~1 we show the measured values of the fractal 
dimensions for $D = 10$ and node decimation
$k=0$, 1, and 2.  For ten scalar fields coupled
to dynamical triangulations, the geometry is expected to
degenerate into branched polymers with $d_H = 2$.  
This is in reasonable agreement
with what we observe, $d_H \approx 2.3$, especially as we use
triangulations of relatively modest size\footnote{The Hausdorff
dimension usually requires much larger triangulations
for its accurate determination, especially if its value 
is large.  This prevented us from measuring $d_H$ for
a single scalar field, where numerical simulations 
indicate $d_H \approx 4$ \cite{kost}.}.  The important observation
is that the estimate of $d_H$ does not change notably as
the node decimation is iterated.

\setlength{\tabcolsep}{20pt}
\begin{table}[t]
 \begin{center}
 \caption{\small The fractal dimensions $d_H$, for ten
  scalar fields coupled to dynamical triangulations,
  after $k=0$, 1, and 2 node decimations.}
 \vspace{10pt}
 \label{tab:hausd}
 \begin{tabular}{|cc|}
 \hline 
   $k$ & $d_H$ \\  \hline
   0 & 2.36(9)  \\
   1 & 2.29(7)  \\
   2 & 2.28(7)  \\ \hline
 \end{tabular}
 \end{center}
\end{table}

The string susceptibility exponent $\gamma_s$ defines the
leading singular behavior of the grand-canonical partition
function:
$Z(\mu) \approx Z_{\rm reg} + (\mu - \mu_c)^{2-\gamma_s}$.
This in turn implies that the canonical partition
function behaves asymptotically as:
$Z(N) \sim \exp (\mu_c N)\,N^{\gamma_s-3}$.
A powerful method to measure $\gamma_s$ is provided by the 
distribution of {\it minimal neck baby universes} ---
a part of the triangulation connected to the rest
{\it via} a minimal neck consisting of three links.  
The distribution of baby universes of size $B$ on a triangulation 
of total volume $N$ can be written as \cite{JainMath92}:
\begin{equation}
 \label{distrib}
 b_N(B) \;\approx\; \frac{B \; Z(B)\;(N-B)\; Z(N-B)}{Z(N)} 
 \;\sim\; (N-B)^{\gamma_s-2}B^{\gamma_s-2}.
\end{equation}
By measuring the distribution $b_N(B)$ on a fixed volume,
$\gamma_s$ is determined by a fit to Eq.~(\ref{distrib})
\cite{AmbJain93}.

\begin{figure}
 \epsfxsize=4in \centerline{\epsfbox{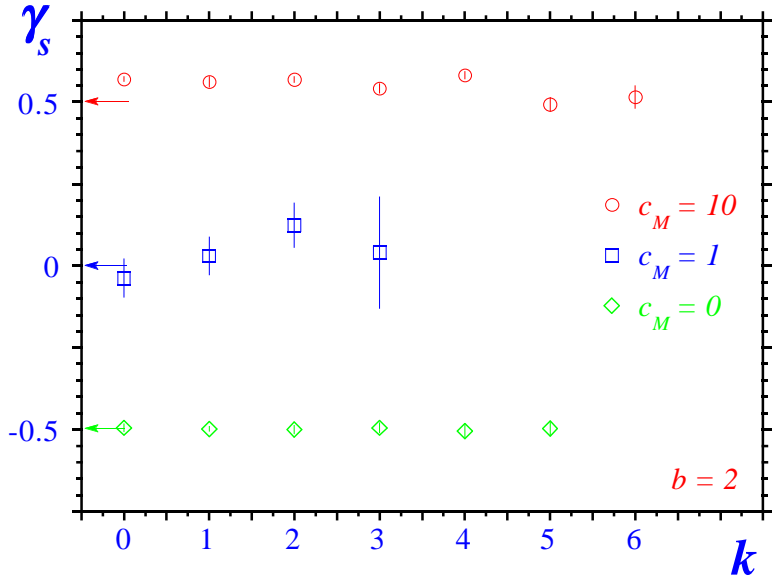}}
 \caption{\small Measured values of the string susceptibility 
   exponent $\gamma_s$ {\it versus} the number of
   node decimations $k$, for zero, one and 
   ten scalar fields coupled to $2d$--gravity. }
 \label{gamma_s}
\end{figure}

We measured $\gamma_s$ for $D=0$, 1, and 10,
and for up to six iterations of the node decimation\footnote{For
one scalar field the determination of $\gamma_s$
is complicated by logarithmic corrections to the free energy
Eq.~(\ref{partition}).
This requires a fit to a more complicated functional form
than Eq.~(\ref{distrib}), which in turn reduces the
precision with which $\gamma_s$ is determined 
\cite{AmbJain93}.}.  The result is shown in Figure~\ref{gamma_s}. 
And, just as for the fractal dimension, the values
of $\gamma_s$ proved to be remarkable stable as the 
blocking procedure is iterated, in some cases even after as 
much as 98\% of the triangulation has been removed ($k=6$).  
Combined, the measurements of $d_H$ and $\gamma_s$
provide strong evidence that the fractal
structure is indeed preserved under this blocking scheme
for both the models we consider.

\subsection{\it Stability Matrix for the Matter Sector}

We now turn to the blocking of the matter fields following 
the prescription in Section~3.2.  We calculate the
stability matrix $T_{\alpha \beta}(\phi,\phi^{\prime})$
using the basis of operators Eq.~(\ref{basis}).
In principle, on a finite lattice operators of order
$m \lesssim N^{1/d_H}$ could be included in the 
analysis; in practice, though, working with a
large basis of operators is troublesome as it
frequently leads to the appearance of complex 
(unphysical) eigenvalues.
This necessitates a truncation of the basis;  
we found that including four operators provided 
sufficient overlap with the relevant operators in 
the fixed point action.
We use a bare lattice of volume $N^{(0)} = 4000$, and
a sequence of blocked lattices:
$N^{(k)} = 2000$, 1000, 500, and 250,
produced by $k=1,2,3,$ and, 4 node decimations, respectively.
The field blocking was repeated several times using different
sets of parameters, $a_w$ and $\xi$, to determine
their optimal values.    

The simulations of one scalar field, $c_M = 1$, served as somewhat 
of a test case as analytical results are available. 
We found the leading and sub-leading eigenvalues, 
$\lambda_1$ and $\lambda_2$, 
to be remarkably insensitive to the choice of $a_w$, as long
as $a_w > 50$. This is demonstrated in Table~2.
This lower limit is understandable as for values of $a_w < 50$ 
the noise term in Eq.~(\ref{fieldblock}) becomes comparable in 
magnitude to the fields themselves. 
It is, on the other hand, rather surprising that suppressing the
noise altogether, choosing $a_w \gg 50$, yields equally stable 
eigenvalues.  Naively, one expects that a fully 
deterministic blocking, $a_w = \infty$, should lead to a
less local fixed point action \cite{BellWils75}.
However, contrary to blocking scalar
fields on a regular lattice, the randomness in blocking the
geometry   
may introduce sufficient noise to limit the range of the 
interactions in the fixed point action.

\setlength{\tabcolsep}{5pt}
\begin{table}
 \caption{\small The two largest eigenvalues of 
   the stability matrix, $\lambda_1$
   and $\lambda_2$, for one
   scalar field ($c_M=1$), for different values of
   the parameter $a_w$, after up to four RG iterations $k$.  
   A field renormaliztion $\xi=1.00$ and
   basis of four field operators ${\cal O}_m$ is
   used in calculating $T_{\alpha \beta}$. }
 \label{tab.cm1A}
 \begin{center}
  {\small
  \begin{tabular}{|r|llll|llll|} \hline
        &  \multicolumn{4}{|c}{$\lambda_1$}
	&  \multicolumn{4}{|c|}{$\lambda_2$} \\ 
   $a_w$ & $k= 1$ & $k=2$ & $k=3$ & $k=4$ 
         & $k= 1$ & $k=2$ & $k=3$ & $k=4$ \\ \hline
  4     & 1.976(6) & 1.961(16) & 1.944(13) & 1.927(25) 
        &  0.11(1) & 0.24(5)   & 0.39(5)   & 0.40(4) \\
  10    & 1.991(8) & 2.002(10) & 1.964(13) & 1.987(19)
        &  0.38(2) & 0.49(4)   & 0.55(3)   & 0.55(4) \\ 
  24    & 1.991(7) & 1.994(8)  & 2.016(10) & 1.980(16)
        &  0.70(2) & 0.78(3)   & 0.80(3)   & 0.81(2) \\
  32    & 2.001(5) & 1.991(9)  & 1.995(8)  & 2.001(12)
        &  0.81(3) & 0.81(3)   & 0.83(3)   & 0.83(3) \\ 
  70    & 1.992(5) & 1.988(8)  & 1.971(13) & 1.974(15)
        &  0.91(5) & 0.97(3)   & 0.94(3)   & 1.02(3) \\ 
  100   & 1.995(6) & 2.001(8)  & 1.980(8)  & 1.975(14) 
        &  0.94(4) & 1.02(3)   & 1.00(3)   & 1.00(2) \\
  130   & 1.992(6) & 1.998(5)  & 1.991(10) & 1.980(19) 
        &  0.97(5) & 0.99(4)   & 1.02(3)   & 1.00(2) \\
  200   & 1.992(5) & 1.992(7)  & 2.005(8)  & 1.997(16) 
        &  1.01(4) & 1.04(3)   & 1.02(3)   & 1.03(3) \\
  400   & 1.998(7) & 1.984(9)  & 2.002(9)  & 1.973(11) 
        &  0.99(4) & 1.04(3)   & 1.04(2)   & 1.07(3) \\
  1000  & 1.992(5) & 2.002(6)  & 1.976(8)  & 1.964(20)
        &  1.05(4) & 1.02(3)   & 1.00(3)   & 1.05(3) \\ 
  10000 & 1.986(5) & 1.999(8)  & 1.998(8)  & 1.991(15)
        &  1.03(5) & 1.02(4)   & 1.08(3)   & 1.02(2) \\ 
  40000 & 1.990(5) & 1.999(6)  & 2.001(9)  & 1.989(15)
        &  0.99(7) & 0.98(5)   & 1.04(4)   & 1.05(3) \\ \hline
  \end{tabular}
  }
 \end{center}
\end{table}

The two leading eigenvalues do, on the other hand, depend strongly on
the field renormalization constant $\xi$.  This we show 
in Table~3 for different values of $\xi$ and using, as
a reasonable amount of noise, $a_w = 100$.  
Using the criteria that there should
exist a (stable) marginal sub-leading eigenvalue, $\lambda_2$,
we judge the optimal value to be $\xi^*=1.00(3)$. 
The quoted error indicates the interval where $\lambda_2$ differs
from unity by one standard deviations.

The combination, $\xi^*=1$ and $a_w^*=100$, yields a 
leading eigenvalue $\lambda_1=1.980(8)(90)(10)$.  
The numbers in the parentheses give the statistical error
on $\lambda_1$, and the errors due to the uncertainty in 
$\xi$ and $a_w$, respectively.
The stability of the two leading eigenvalues under repeated
RG iterations is shown in Figure~4.

\begin{figure}
 \epsfxsize=4in \centerline{\epsfbox{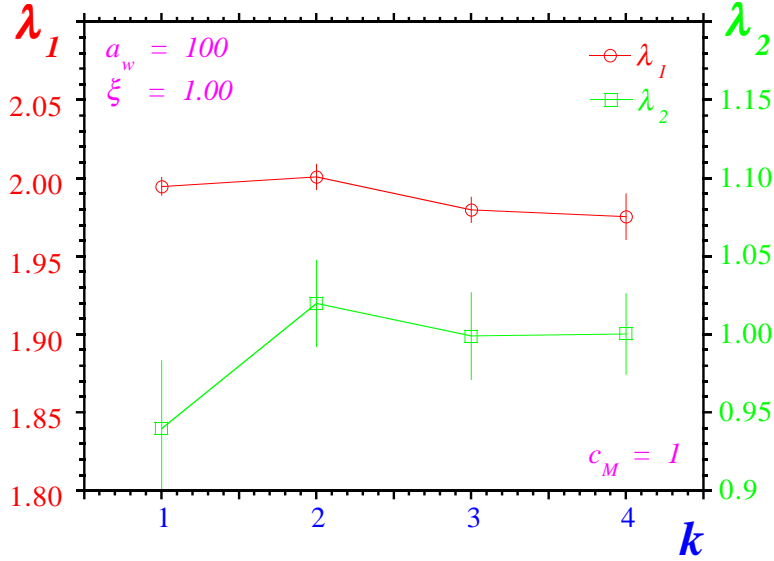}}
 \label{fig.cm1}
 \caption{\small The two leading eigenvalues of the stability matrix for
   one scalar field ($c_M = 1$), calculated using $a_w=100$ 
   and $\xi=1.00$, {\it versus} the number of RG iterations $k$.} 
\end{figure}

\begin{table}
 \caption{\small Same as in Table~\ref{tab.cm1A}, except
  for varying field renormalization $\xi$ but 
  fixed amount of noise $a_w = 100$.}
 \label{tab.cm1B}
 \begin{center}
  {\small
  \begin{tabular}{|r|llll|llll|} \hline
         &  \multicolumn{4}{|c}{$\lambda_1$}
	 &  \multicolumn{4}{|c|}{$\lambda_2$} \\  
   $\xi$ & $k= 1$   & $k=2$     & $k=3$     & $k=4$ 
         & $k= 1$   & $k=2$     & $k=3$     & $k=4$ \\ \hline  
   0.92  & 2.360(8) & 2.350(10) & 2.361(10) & 2.340(22)
         & 1.09(6)  & 1.14(4)   & 1.14(3)   & 1.11(4) \\ 
   0.96  & 2.175(6) & 2.166(10) & 2.156(10) & 2.148(20) 
         & 0.94(6)  & 1.09(4)   & 1.07(2)   & 1.08(2) \\ 
   0.98  & 2.085(6) & 2.076(6)  & 2.078(13) & 2.064(18)
         & 0.97(5)  & 1.06(3)   & 1.01(2)   & 1.04(3) \\
   1.00  & 1.995(6) & 2.001(8)  & 1.980(8)  & 1.975(14) 
         & 0.94(4)  & 1.02(3)   & 1.00(3)   & 1.00(2) \\
   1.02  & 1.925(8) & 1.907(6)  & 1.900(7)  & 1.903(13)
         & 0.87(5)  & 1.00(4)   & 0.97(3)   & 0.99(2) \\ 
   1.04  & 1.855(6) & 1.848(4)  & 1.845(10) & 1.837(13)
         & 0.89(4)  & 0.93(3)   & 0.94(1)   & 0.95(2) \\
   1.08  & 1.715(4) & 1.705(5)  & 1.710 (8) & 1.700(9)
         & 0.84(4)  & 0.90(3)   & 0.89(2)   & 0.86(3) \\ \hline
  \end{tabular}
  }
 \end{center}
\end{table}

\setlength{\tabcolsep}{3.5pt}
\begin{table}
 \caption{\small Same as in Table~\ref{tab.cm1A}, except for
  ten scalar fields, $c_M =10$, and a different field
  renormalization, $\xi=0.92$. The asterisk indicates a
  complex eigenvalue.}
 \label{tab.cm10A}
 \begin{center}
  {\small
  \begin{tabular}{|r|llll|llll|} \hline 
         &  \multicolumn{4}{|c}{$\lambda_1$}
	 &  \multicolumn{4}{|c|}{$\lambda_2$} \\
   $a_w$ & $k= 1$        & $k=2$         & $k=3$      & $k=4$ 
         & $k= 1$        & $k=2$         & $k=3$      & $k=4$     \\ \hline
   4     & 2.361(4)      & 2.339(8)      & 2.323(17)  & 2.290(21)
         & 0.041(12)     & 1.27(11)      & 1.12(11)   & 1.07(14)  \\
   24    & 2.366(3)      & 2.360(6)      & 2.345(7)   & 2.337(10) 
         & 0.32(5)       & 1.23(10)      & 1.22(11)   & 0.95(9)   \\
   32    & 2.361(4)      & 2.347(5)      & 2.355(7)   & 2.350(10)
         & 0.46(48)      & 1.41(27)      & 1.02(18)   & 1.07(17)  \\
   70    & 2.353(4)      & 2.352(6)      & 2.350(6)   & 2.352(11)
         & 0.64(7)       & 1.30(23)      & 0.95(11)$\ast$ & 1.02(4)   \\
   100   & 2.360(3)      & 2.359(5)      & 2.359(9)   & 2.332(10)
         & 0.63(3)$\ast$ & 1.05(7)$\ast$ & 1.01(5)    & 1.00(3)   \\
   130   & 2.360(4)      & 2.351(5)      & 2.362(8)   & 2.331(10)
         & 0.69(3)$\ast$ & 1.07(9)       & 0.95(5)    & 1.08(3)   \\
   200   & 2.363(4)      & 2.357(6)      & 2.334(6)   & 2.344(14)
         & 0.73(5)$\ast$ & 1.02(5)       & 1.07(4)    & 1.09(3)   \\
   400   & 2.356(3)      & 2.358(6)      & 2.359(8)   & 2.349(8)
         & 0.92(5)$\ast$ & 0.96(5)       & 0.99(4)    & 1.06(3)   \\
   1000  & 2.351(4)      & 2.354(4)      & 2.351(10)  & 2.339(11)
         & 0.94(6)$\ast$ & 0.87(8)$\ast$ & 1.01(3)    & 1.09(3)   \\
   10000 & 2.365(4)      & 2.352(5)      & 2.351(6)   & 2.341(13)
         & 0.87(8)$\ast$ & 0.94(6)       & 1.02(4)    & 0.93(8)   \\
$\infty$ & 2.354(4)      & 2.360(6)      & 2.356(8)   & 2.334(11)
         & 1.00(7)$\ast$ & 1.01(8)       & 1.05(3)    & 1.01(4)   \\ \hline
  \end{tabular}
  }
 \end{center}
\end{table}

\setlength{\tabcolsep}{4pt}
\begin{table}
 \caption{\small Same as in Table~\ref{tab.cm10A}, except
  for variable field renormalization $\xi$ and using 
  fixed amount of noise, $a_w = 100$.}
 \label{tab.cm10B}
 \begin{center}
  {\small
  \begin{tabular}{|r|llll|llll|} \hline 
         &  \multicolumn{4}{|c}{$\lambda_1$}
	 &  \multicolumn{4}{|c|}{$\lambda_2$} \\  
   $\xi$ & $k= 1$         & $k=2$          & $k=3$ & $k=4$ 
         & $k= 1$         & $k=2$          & $k=3$ & $k=4$ \\ \hline 
   0.84  & 2.823(4)       & 2.831(9)       & 2.813(11)     & 2.790(10)
         & 0.62(5)$\ast$  & 1.09(4)$\ast$  & 1.10(12)      & 1.12(4)   \\
   0.88  & 2.578(3)       & 2.581(5)       & 2.582(6)      & 2.567(11)
         & 0.76(12)       & 0.99(8)        & 1.09(7)       & 1.06(5)   \\
   0.90  & 2.467(4)       & 2.463(6)       & 2.463(9)      & 2.426(13)
         & 0.72(9)        & 1.06(11)       & 1.08(6)       & 1.13(5)   \\ 
   0.92  & 2.360(3)       & 2.359(6)       & 2.359(9)      & 2.332(10)
         & 0.63(3)$\ast$  & 1.07(5)$\ast$) & 1.01(5)       & 1.00(3)   \\
   0.94  & 2.169(3)       & 2.179(6)       & 2.164(6)      & 2.162(12)
         & 0.62(11)$\ast$ & 1.16(8)        & 1.00(8)       & 0.94(4)   \\
   0.96  & 2.169(3)       & 2.179(6)       & 2.164(6)      & 2.162(12) 
         & 0.53(4)$\ast$  & 1.21(30)       & 1.00(6)       & 0.94(3)   \\
   1.00  & 1.993(3)       & 1.996(5)       & 1.986(7)      & 1.990(8)
         & 0.58(3)$\ast$  & 0.91(9)        & 0.83(6)$\ast$ & 0.87(4)   \\
   \hline
  \end{tabular}
  }
 \end{center}
\end{table}

\begin{figure}
 \epsfxsize=4in \centerline{\epsfbox{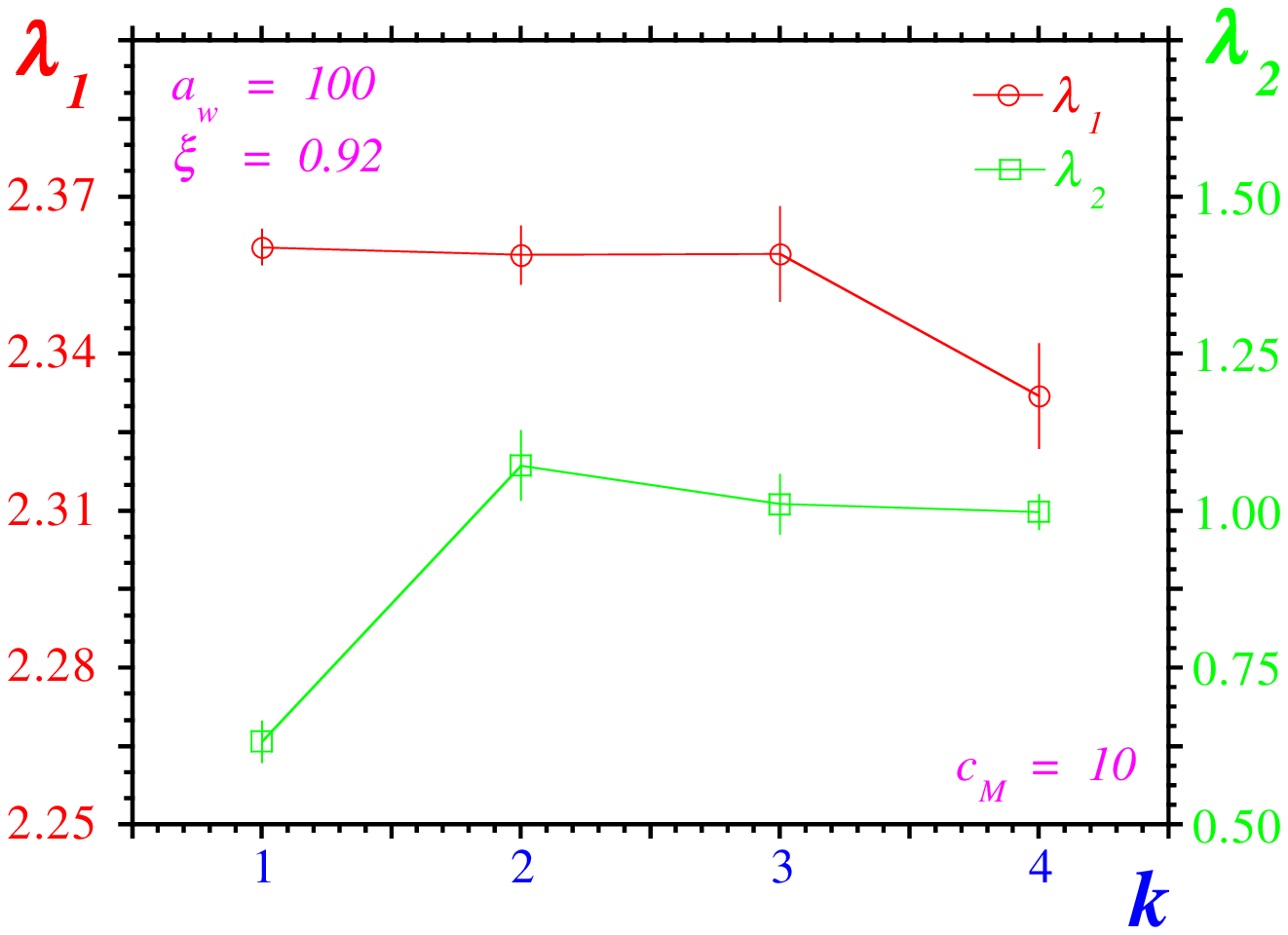}}
 \label{fig.cm10}
 \caption{\small The two leading eigenvalues of the stability matrix for
   ten scalar fields ($c_M = 10$), using $a_w=100$ and $\xi=0.92$,
   {\it versus} the number of RG iterations $k$.} 
\end{figure}

We studied the model with ten scalar fields using the same procedure.  
The stability under blocking of the leading and sub-leading eigenvalues 
is shown for different amount of noise $a_w$ in Table~4. 
We select the value $a_w=100$ as the most optimal, though 
again, values of  $a_w > 100$ all seem to produce 
relatively stable eigenvalues.  In contrast to the $c_M = 1$
model, however, for $c_M = 10$ even using a bases of 
only four field operators leads occasionally
complex and unphysical eigenvalues.  This appears usually
in the first iteration of the field blocking and for the
sub-leading eigenvalue $\lambda_2$.  
We took this as indicating that the bare system is
not sufficiently close to the critical fixed point for 
the linear approximation Eq.~(\ref{expand2}) to be valid; 
this observation is supported by the large change in the
eigenvalues in the first iteration.
This is not terribly troubling, though, as one can choose $a_w$ such
that in the successive blocking all of the eigenvalues are real.

In Table~5 we show the variation in the eigenvalues
with the field renormalization.
Using the same analysis as we did for $c_M=1$, we find that 
$\xi=0.92(4)$ gives a sub-leading eigenvalue close to unity 
for $c_M=10$.  This value, combined with $a_w = 100$,
corresponds to a leading eigenvalue $\lambda_1=2.36(20)$. 
Here we just quote the dominant error due to the uncertainty 
in determining $\xi^*$. 
In Figure~5 we show the stability of the two largest 
eigenvalues as the fields are blocked 
using $\xi^* = 0.92$ and $a_w^* = 100$.

\begin{figure}
 \epsfxsize=4in \centerline{\epsfbox{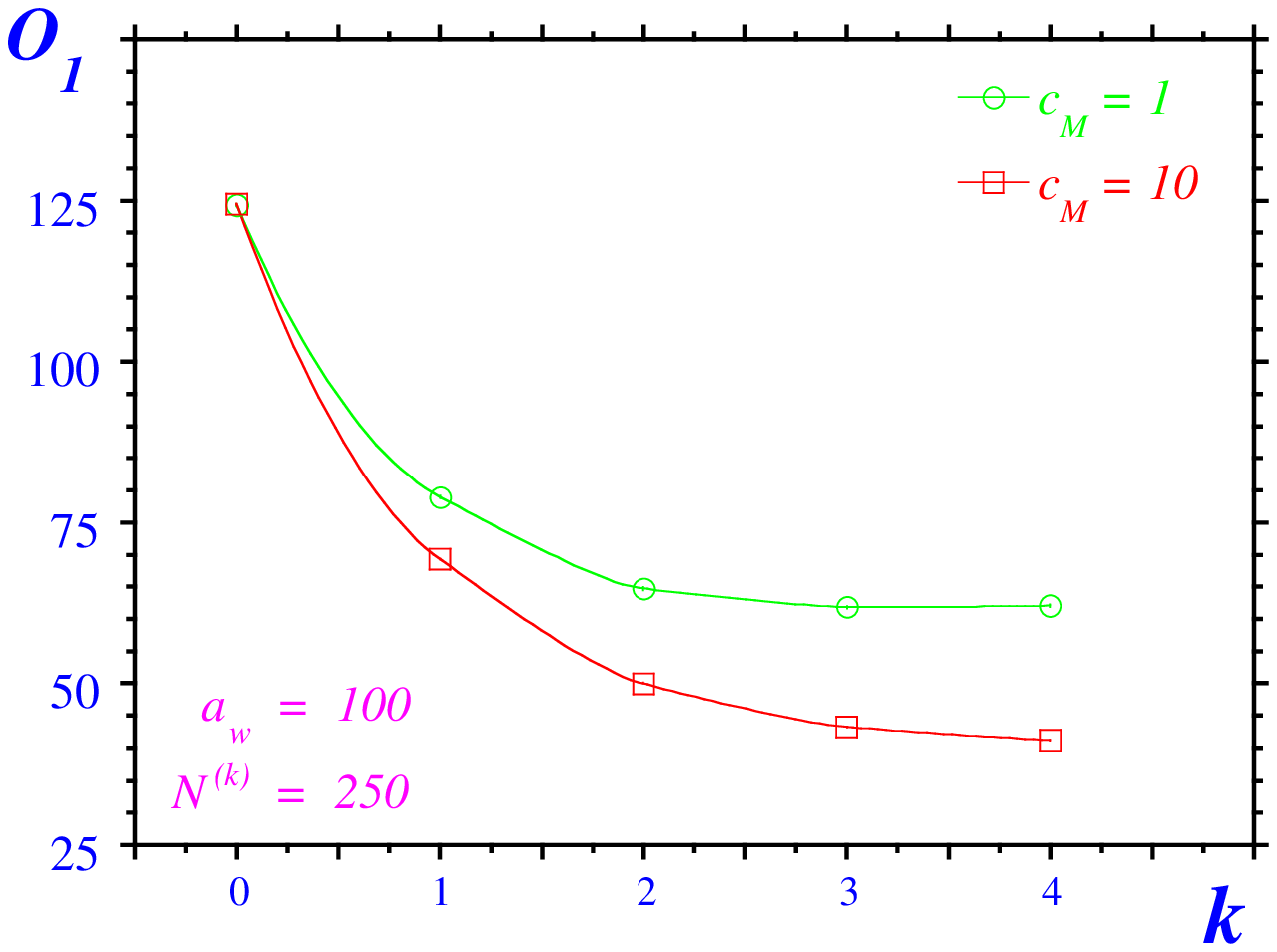}}
 \label{fig.flow}
 \caption{\small The ``flow'' of the (normalized) operator 
   ${\cal O}_1 = \langle \phi_i \square \phi_i \rangle$ for both one
   and ten scalar fields.  All the measurements are done on 
   triangulations of volume 250, generated by different amount
   $k$ of node decimation from bare lattices of
   volume $2^{k} 250$.  The fields are blocked using
   $a_w = 100$, and $\xi = 1.00$ and 0.92 for $c=1$ and 10 
   respectively.} 
\end{figure}

Additional evidence for a fixed point structure is provided by
the flow of the operators, Eq.~(\ref{basis}), as the field
blocking is iterated.  This is demonstrated in Figure~6
for the expectation value of the operator 
${\cal O}_1 = \langle \phi_i \square \phi_i \rangle$, 
both for $c_M=1$ and 10.  
Notice that in order to eliminate undue influence 
of geometric finite-size effects, 
we only compare operators measured at the {\it same} 
blocked lattice volume, $N^{(k)}=250$, but which corresponding
to different amount of node decimation.
For both models, the value of the
operator changes substantially in the first blocking
iteration; this indicates that bare model 
is relatively far away from the critical fixed point 
This effect is more pronounced for the $c_M = 10$ model,
as discussed earlier.  However, already after the
second blocking of the fields, the operators appear 
to have converged to their fixed point values.

\section{Discussion}

Although there exists no analytic RG calculation for dynamical lattices, 
using the results of matrix model calculations one
can nevertheless predict the appropriate value of $\xi$, 
the field renormalization constant, for the case $c_M=1$.  
Additionally, simple dimensional arguments yield an
estimate of some of the eigenvalues of the stability matrix.

First lets look at the origin of the length dimension in 
the lattice fields.  On the bare lattice we simulate a system with 
fields governed by the continuum action 
(let us take the internal dimension $d$ to be arbitrary for the moment),
\begin{equation}
\label{cont_action}
 S_{\mbox{cont}} \;=\;
 \int {\rm d}^dx \sqrt{g} \; 
 \left ( \partial \phi_{\mbox{cont}}(x) \right )^2.
\end{equation}
If we scatter points separated by a distance $h$ throughout the 
manifold, we can approximate this action by
\begin{equation}
  S_{\mbox{cont}} \;\approx\; \sum_{<i,j>} 
 \frac{1}{h^2}
 \; \left ( \phi_{\mbox{cont}}(x_i)-
  \phi_{\mbox{cont}}(x_j)\right ) ^2h^d.
\end{equation}
To write this in terms of lattice fields, as we did in 
Eq.~({\ref{action}), requires
\begin{equation}
 \phi_{\mbox{cont}} \;=\; \phi_{\mbox{lat}} h^{\frac{2-d}{2}}.
\end{equation}
The continuum field, $\phi_{\mbox{cont}}$, has length 
dimension $\beta=-\eta/2$ defined by the scaling of the 
geodesic two-point function,
\begin{equation}
 \langle \phi_{\mbox{cont}}(x_i) \; \phi_{\mbox{cont}}(x_j) \rangle 
  \;\sim\; r_{ij}^{-\eta}.
\end{equation}
This implies that a change of the length scale of problem, 
$h \rightarrow sh$, should induce a change of scale of the 
(continuum) fields:
\begin{equation}
 \phi_{\mbox{cont}} \;\rightarrow\; \phi_{\mbox{cont}}s^{-\beta}.
\end{equation}
For this to occur, the lattice fields must re-scale under such 
a change like
\begin{equation}
 \phi_{\mbox{lat}} \;\rightarrow\; 
 \phi_{\mbox{lat}}s^{\frac{d-2}{2}-\beta}.
\end{equation}

In our lattice simulation we do not have direct access to the length scale, 
but we do have access to the volume scale. So on a lattice with 
fractal dimension 
$d_H$, a volume rescaling of $b=s^{d_H}$ requires the fields be rescaled like
\begin{equation}
 \label{philatres}
 \phi_{\mbox{lat}} \;\rightarrow\; 
 \phi_{\mbox{lat}}b^{\frac{d-2-2\beta}{2d_H}},
\end{equation}
The implication then, is that in $d=2$ the field renormalization constant
\begin{equation}
 \label{scale_exp}
 \xi \;=\; b^{-\frac{\beta}{d_H}},
\end{equation}
This is a generalization of the field renormalization constant given in 
Ref.~\cite{BellWils74}.

Consider first the case $c_M=1$. 
If we assume the action to be  dimensionless, then we can count dimensions
to show that the undressed length dimension for a Gaussian field in a flat
two-dimensional space is zero:
\begin{equation}
 S \;=\; \int d^2x(\partial \phi)^2  
 \;\;\Longrightarrow \;\; \beta_0 = 0.
\end{equation}
In the presence of quantum gravity, however, the field will in general develop
an anomalous length dimension.  For $c_M\leq1$ we can use the well known 
KPZ \cite{KPZDDKref} scaling relations:
\begin{equation}
\label{KPZ1}
\beta - \frac{\beta (1-\beta)}{1-\gamma_s} \;=\; \beta_0,
\end{equation}
with
\begin{equation}
\label{KPZ2}
 \gamma_s \;=\; \frac{1}{12}\left ( c_M-1-\sqrt{(25-c_M)(1-c_M)}\right ),
\end{equation}
to find $\beta$, the dressed field dimension.
So for $c_M=1$, 
\begin{equation}
 \label{beta_result}
 \beta \;=\; \beta_0 \;=\; 0,
\end{equation}
implying $\xi = 1$.  For $c_M=1$ our MCRG result of $\xi^*=1.00(3)$ can be taken as 
an independent numerical determination of $\beta$ consistent with 
Eq.~(\ref{beta_result}). For $c_M=10$, our result of $\xi^*=0.92(4)$ 
would suggest that $\frac{\beta}{d_H}=0.12(6)$.

Although meaningful analytical results are not available for $c_M > 1$,
simple dimensional arguments showing consistency of our 
$c_M=10$ results. Consider the operator $O_0=\langle \sum_i \phi_i^2 \rangle$.
From the analysis of the eigenvectors of the
stability matrix this is essentially the operator corresponding
to the leading eigenvalue. The simplest ansatz for its
behavior under scaling follows from the scaling of the field in 
Eq.~(\ref{philatres})
$$O_0 \;\to\; b^{-\frac{2\beta}{d_H}-1}O_0\;.$$
The corresponding coupling 
constant $g_{O_0}$
in a dimensionless action should re-scale inversely, like 
\begin{equation}
 g_{O_0} \;\rightarrow\; 
 b^{1+\frac{2\beta}{d_H}} g_{O_0},
\end{equation}
suggesting that the leading relevant eigenvalue should be
\begin{equation}
 \lambda_{O_0} \;=\; b^{1+\frac{2\beta}{d_H}}.
\end{equation}
The extra factor of $b$ is due to the change of the sum over $N$ nodes to a
sum over $N'=N/b$ nodes.

\begin{figure}
 \epsfxsize=4in \centerline{\epsfbox{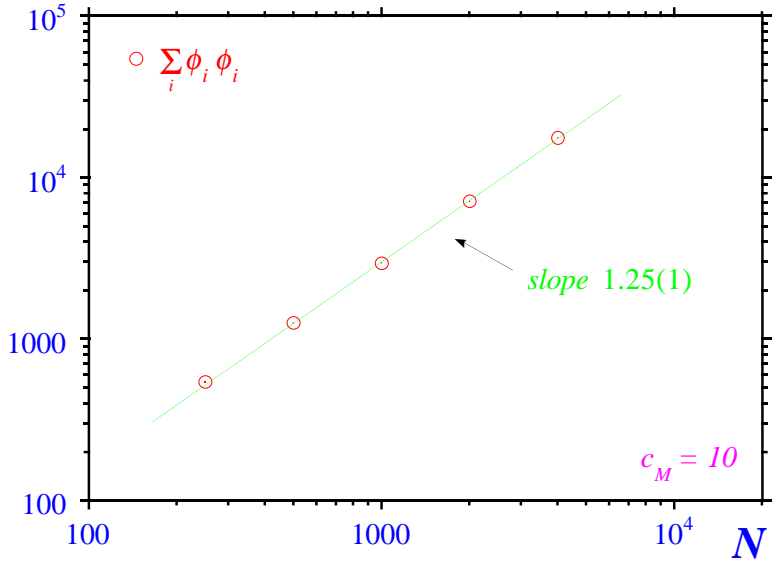}}
 \label{fig.op0}
 \caption{\small The volume scaling of the operator 
  ${\cal O}_0 = \langle\sum_i \phi_i \phi_i\rangle$
  measured on bare triangulations with $c_M=10$.  The 
  straight line is a fit 
  $\langle\sum_i \phi_i \phi_i\rangle \sim V^{1.25(1)}$.} 
\end{figure}

We note that substituting $\beta/d_H=0$, for $c_M=1$, and $\beta/d_H=0.12$, 
for $c_M=10$, gives $\lambda_{O_0}=1.98(10)$ and 
$\lambda_{O_0}=2.36(20)$, respectively.  This is remarkably
consistent with the leading eigenvalues we obtained for both cases.

The remaining check that we can institute is that the critical exponents 
extracted from the eigenvalues of the stability matrix are consistent with 
those obtained through finite-size scaling analysis.
Simple arguments show that if an operator $O$ at
criticality scales like 
\begin{equation}
 \langle O \rangle \;\sim\;  N^{\mu} 
\label{rgscale}
\end{equation}
on lattices of volume $N$ then there should be a leading eigenvalue
\begin{equation}
 \lambda_1 \;=\; b^{\,\mu}.
 \label{lead_eiv}
\end{equation}
Consider the leading operator $O_0$, again for $c_M=10$. 
Its finite-size scaling is shown in Figure~7 and 
yields an exponent $\mu=1.25(1)$. This is consistent with
the leading eigenvalue $\lambda_1=2.36(20)$ extracted from the stability 
matrix (Figure~\ref{fig.cm10}).

\section{Conclusions}

We have successfully extended the renormalization group scheme for
dynamical triangulations based on node decimation to the case of
continuous scalar fields coupled to two-dimensional
gravity. Results for the wave function renormalization
constant $\xi$ and hence $\frac{\eta}{d_H}$ together with
the leading eigenvalue of the stability matrix are obtained for $c_M=1$ and
$c_M=10$ which are consistent with theoretical expectations.


\begin{thebibliography}{99}

{\raggedright

\bibitem{KPZDDKref}
  F.~David, {\it Mod.~Phys.~Lett.~B} {\bf 186} (1987) 379; \\
  J.~Distler and H.~Kawai, {\it Nucl.~Phys.~B} {\bf 321} (1989) 509; \\
  V.G~Knizhnik, A.M.~Polyakov and A.~Zamolodchikov, 
   {\it Mod.~Phys.~Lett.~A} {\bf 3} (1988) 819.

\bibitem{ThorCatt96}
 G.~Thorleifsson and S.M.~Catterall, 
  {\it Nucl.~Phys.~B} {\bf 461} (1996) 350.

\bibitem{neal}   
 R.M.~Neal, {\it Suppressing Random Walks in Markov Chain Monte Carlo
  Using Ordered Overrelaxation}, Technical report No.~9508, Dept.\ of
  Statistics, University of Toronto ({\tt bayes-an/9506004}). 

\bibitem{kennedy}
 I.~Horv\'{a}th and A.D.~Kennedy, 
  {\it Nucl.~Phys.~B} {\bf 510} (1998) 367.
 
\bibitem{Lang86}
 C.B.~Lang, {\it Nucl.~Phys.~B} {\bf 265} (1986) 630.

\bibitem{BellWils75}
  T.L.~Bell and K.G.~Wilson, {\it Phys.~Rev.~B} {\bf 11} (1975) 3431.

\bibitem{CattThor95}
  S.~Catterall, G.~Thorleifsson, M.~Bowick and V.~John, 
   {\it Phys.~Lett.~B} {\bf 354} (1995) 58.

\bibitem{AmbJur95}
  J.~Ambj\o rn, J.~Jurkiewicz and Y.~Watabiki, 
   {\it Nucl.~Phys.~B} {\bf 454} (1995) 313.
   
\bibitem{kost}
  J.~Ambj\ orn and K.N.~Anagnostopoulos,
   {\it Nucl.~Phys.~B} {\bf 497} (1997) 445.

\bibitem{JainMath92} 
  S.~Jain and S.D.~Mathur, 
   {\it Phys.~Lett.~B} {\bf 286} (1992) 239.

\bibitem{AmbJain93}
  J.~Ambj\o rn, S.~Jain, and G.~Thorleifsson, 
   {\it Phys.~Lett.~B} {\bf 307} (1993) 34; \\
  J.~Ambj\o rn, and G.~Thorleifsson, 
   {\it Phys.~Lett.~B} {\bf 323} (1994) 7.

\bibitem{David92}
 F. David, {\it Simplicial Quantum Gravity and Random Lattices},
 ({\tt hep-th/9303127}), Lectures given at Les Houches Summer School 
  on Gravitation and Quantization, Session LVII, 
  Les Houches, France, 1992; \\
 J. Ambj\o rn, {\it Quantization of Geometry}, ({\tt hep-th/9411179}), 
  Lectures given at Les Houches Summer School on Fluctuating 
  Geometries and Statistical Mechanics, Session LXII, 
  Les Houches, France 1994;  \\
 P. Di Francesco, P. Ginzparg and J. Zinn-Justin, 
  {\it Phys.~Rep.\ } {\bf 254} (1995) 1.
  
\bibitem{Bennet88}
  A.~Bennett, {\it Nucl.~Phys.~B} {\bf 300} (1988) 253.

\bibitem{BellWils74}
  T.L.~Bell and K.G.~Wilson, {\it Phys.~Rev.~B} {\bf 10} (1974) 3935.
}

\end{thebibliography}
\end{document}